\documentclass[10pt]{article}
\usepackage{graphicx}
\usepackage{color}

\begin{document}

\begin{titlepage}
\begin{flushright}
TU- 709\\
\end{flushright}
\ \\
\begin{center}
\LARGE
{\bf
 Complete Resolution \\
   of the Quantum Zeno Paradox\\
   for Outside Observers
}
\end{center}
\ \\
\begin{center}
\Large{
M.Hotta${}^\ast$ and M.Morikawa${}^\dagger$    }\\
{\it
${}^\ast$
Department of Physics, Faculty of Science, Tohoku University,\\
Sendai 980-8578,Japan\\
hotta@tuhep.phys.tohoku.ac.jp \\

\ \\

${}^\dagger$
Physics Department, Ochanomizu University,\\
Tokyo, 112-8610, Japan \\
hiro@phys.ocha.ac.jp
}
\end{center}

\begin{abstract}
The quantum Zeno paradox is fully resolved for  
purely indirect and incomplete measurements performed 
by the detectors outside the system.
If the outside detectors are prepared to observe propagating signals of 
 a decay event of an excited state in the core region, 
 the survival probability of the state is not changed at all 
  by the outside measurements, 
 as long as the wavefunction of the signals does not have 
 reflectional-wave contributions going back to the core 
  by the outside detectors. The proof is independendent of the decay law 
 of the survival probabilities.
Just watching frequently from outside (observation) cannot be regarded 
as a measurement which yields the quantum Zeno effect.  

\end{abstract}

\end{titlepage}

\section{Introduction}

The quantum Zeno (QZ) paradox \cite{misra77} has been tempting a lot of
physicists to make the complete and elegant resolution for long years. The
quantum theory predicts, as will be mentioned in the next section, that
quantum processes are frozen by their continuous measurements. \ However the
predicted freezing of the processes apparently contradicts macroscopic
reality. \ 

To express the QZ paradox microscopically, a phrase is available; \textit{an
unstable particle does not decay under the frequent monitoring}. For
example, the proton was expected to decay within the lifetime of order $%
10^{30}$ years and sensitive experiments for the proton decay have been
performed. Unfortunately, no signal has been found so far and the elementary
particle physicists have changed the ``wrong'' Lagrangians of their models
for the decay so as to prolong the lifetime. \ From the view point of the QZ
effect, there is another possible solution to explain why the proton decay
has not actually been observed\cite{proton}; sensitive devices in the
experiment have monitored the decay process too frequently and therefore
have frozen the decay process.

Amazingly, the QZ paradox is also applied to the well known Schr{\"{o}}%
dinger's cat system. Suppose that an excited state of an atom decays and
emits photons, and that the detection of the photons by a measurement device
triggers the death of his cat. 
Then it is possible to propose an interesting paradox; \textit{the cat's
lifetime is prolonged by the continuous observation of the cat}. We can
assume, once the excited state decays, that the time interval from the
photon emission to the cat's death is instantaneous. Then, each survival
result of the cat by the observation directly implies that the excited state
of the atom has not yet decayed. Thus, the observation of the cat is an
indirect measurement of the atom itself. According to the QZ effect, the
continuous measurement of the atom should prolong the decay of the excited
state. Consequently, the lifetime of the cat should also be prolonged.

Nowadays, the QZ paradox is a quite exciting issue for not only theorists
but also experimentalists. In fact, several experiments for the QZ effect
have been already performed and showed the evidence that the frequent
measurements actually make the quantum processes slower in some systems \cite
{itano90}.

Although such successful experiments have been achieved, we cannot say that
the essential part of the QZ paradox has fully been solved. The experiments
were performed only by use of \textit{direct} measurements of unstable
systems. In order to observe the measurement effect, the system was coupled
directly and strongly with other ancilla states, and(or) was immersed in the
external strong pulsed waves. Clearly, such operations induce additional
interaction terms in the system Hamiltonian and drastically modify the
dynamics in the closed regions. Thus, strictly speaking, there is no
surprise even if the strong \textit{direct} measurements yield prominent
phenomena like freezing of the system evolution. In order to resolve the QZ
paradox completely, the analysis of \textit{purely indirect} measurements of
the system is inevitable \footnote{%
We thank A.Shimizu for stressing the significance of purely indirect
measurements to resolve the QZ paradox. \ }. The QZ paradox becomes indeed
serious, for instance, when the signals emitted by an unstable system at the
decay time propagate away from the system and are measured by outside
devices. However actually, such experiments have not been realized so far.
Therefore we do not have any experimental evidence that the QZ paradox
occurs under the continuous and purely indirect measurements. Note that the
above examples of the QZ paradox are all based on the purely indirect
measurements. \ In the proton decay example, photo-multipliers on the wall
of water storage tank observe Cherenkov beams emitted by electric showers
which are formed by the proton decay in the water. In the Schr{\"{o}}%
dinger's cat example, not the unstable atom but the cat is observed.
Therefore, it is of great importance to check whether the quantum theory
really generates the QZ paradox in the purely indirect measurements or not.

In general, the emitted signals (photons and so on), which imply the decay
of the unstable state, reach the wave zone eventually after the launch from
the core system. We defined naturally in the previous paper \cite{hm} the
wave-zone states and wave-zone subspace for the signals in the Hilbert
space. Then, it was shown that no QZ paradox appears in the indirect and
ideal measurements of the wave-zone states. In this paper, we critically
analyze the QZ paradox again, taking account of finite errors of the general
indirect measurements. Using the definition of the wave-zone states, a no-go
theorem of the QZ paradox under general indirect measurements with finite
errors is proven; no indirect measurements by use of the signals affect the
survival probability of the unstable core state at all when the measurements
of the signals are performed in the wave-zone states, as long as the
wavefunction of the signals do not have any reflectional contributions back
to the core system by the detectors. Thus we argue that the no-go theorem
fully resolves the QZ paradox under purely indirect measurements. For
instance, the quantum theory indeed predicts that the unstable particle
decays even if the outside apparatus monitors continuously, and that the
lifetime of the Schr\"{o}dingers cat is not prolonged by the cat
observations.

The outline of this paper is as follows. In the next section 2, we review
briefly the standard QZ argument. In the section 3, we specify in the
simplest form the general feature of the purely indirect the QZ experiments
with finite errors. Under some conditions, we prove the impossibility of the
QZ effect in the purely indirect distant measurements. Based on this
argument, in section 4, we study a simple model which demonstrates this
impossibility of the QZ effect. In the last section, we summarize the
results.

\section{The Quantum Zeno Paradox under the Continuous and Direct
Measurements}

\ Let us first explain the quantum Zeno (QZ) paradox in the standard way.
Consider a quantum system with unitary evolution. The Hamiltonian is given
by $H$. Suppose that the Hamiltonian $H$ is the sum of the free evolution
term $H_{o}$ and the interaction term $H_{int}$, which generates dynamical
transitions from an eigenstate state $|e\rangle $ of $H_{o}$ to the other
state or decay of the state $|e\rangle $; 
\begin{equation}
H=H_{o}+H_{int},
\end{equation}
where $H_{o}|e\rangle =E_{o}|e\rangle $. Then let us focus on the survival
probability $s(t)$ of the state $|e\rangle $ ; 
\begin{equation}
s(t)=|\langle e(0)|e(t)\rangle |^{2},
\end{equation}
where $|e(t)\rangle $ is the time evolved state from the initial state $%
|e(0)\rangle =|e\rangle $ at time $t$. According to the time reversal
symmetry, $s(t)$ behaves at short times like 
\begin{equation}
s(\Delta t)\sim 1-\alpha \Delta t^{2},
\end{equation}
where $\alpha $ is a positive constant. Therefore, at time $t=N\Delta t$
after N successive \textit{direct} measurements of the state, the survival
probability of the state is evaluated as 
\begin{equation}
s_{N}(t)\sim \left( 1-\alpha (\Delta t)^{2}\right) ^{\frac{t}{\Delta t}}\sim
e^{-\alpha t\Delta t}.
\end{equation}
Consequently in the continuous measurement of the state ( in the limit $%
N\rightarrow \infty $, $\Delta t\rightarrow 0$ with $t$ fixed) one obtains 
\begin{equation}
\lim_{N\rightarrow \infty }s_{N}(t)=1,
\end{equation}
thus the time evolution of the state $|e\rangle $ freezes. This explicitly
shows the appearance of the QZ paradox.

It is known that direct measurements with finite errors can also partially
freeze the time evolution of the system\cite{direct}. Let us consider
ancilla system which represents the measurement apparatus. Suppose that a
direct interaction $gH_{m}$ is turned on between the system and the
measurement apparatus, where $g$ is a real coupling constant and $H_{m}$ is
a Hermitian operator. For simplicity, we assume that the initial state $%
|e\rangle $ is the eigenstate with zero eigenvalue of $H_{m}$; 
\begin{equation}
H_{m}|e\rangle =0.
\end{equation}
The term ``direct'' means here that the measurement Hamiltonian does not
commute with the interaction Hamiltonian $H_{i}$; $[gH_{m},\ H_{i}]\neq 0$.
By adding $gH_{m}$ to $H$, the total Hamiltonian now reads 
\begin{equation}
H_{tot}=H+gH_{m}=H_{o}+H_{i}+gH_{m}.
\end{equation}
The time evolution under the measurement is generated by this $H_{tot}$.
Here finiteness of the coupling $g$ implies incompleteness of the
measurement, that is, the non-ideal measurement with finite errors. Let us
suppose that the coupling $g$ is finite, but large enough and that the
eigensubspace with the zero eigenvalue of the Harmitian operator $H_{m}$ is
non-degenerate, that is, a one-dimensional space. \ Then, it is shown that
the survival probability $s_{g}(t)$ of the state $|e\rangle $ becomes larger
than the value without measurement (the case of $g=0$). This type of
disturbance on the evolution is generally called QZ effect as well. \ 

This general QZ effect under observations with finite errors has a
distinguished property for the measurement of the exponentially decaying
systems. \ The ideal measurements, no matter how frequent, \ do not change
at all the exponential decay law, $s(t)\propto e^{-\Gamma t}$, which appears
in the intermediate time scales of many systems. The following evaluation
shows this fact: 
\begin{equation}
\lim_{N\rightarrow \infty }s_{N}(t)\propto \lim_{N\rightarrow \infty }\left(
1-\Gamma \Delta t\right) ^{\frac{t}{\Delta t}}=e^{-\Gamma t}.
\end{equation}
On the other hand, surprisingly, it is proven \cite{koshino02} that
incomplete measurements can change the exponential decay law and show the QZ
or anti-QZ effect. This fact makes studies of the general measurements with
finite errors more important.

In the next section we study purely indirect measurements with finite errors
and discuss the possibility of the QZ paradox in this case.

\section{No-Go Theorem}

\ In general, the total quantum system including the measurement apparatus
may actually be very complicated. \ However a subspace $\mathcal{H}_{Z}$
which is relevant to consider the QZ effect would be almost closed in
dynamics. We would like to concentrate on this subspace $\mathcal{H}_{Z}$. \
Let $\{|e\rangle ,\{|n\rangle \}_{n=1,2,...}\}$ denote the complete
orthonormal basis vectors in $\mathcal{H}_{Z}$.

Let us denote $|e(t)\rangle $ as the time evolved state at time $t$ from the
initial state $|e(0)\rangle =|e\rangle $. \ Since the evolution of the state
vector is closed in this space, 
\begin{equation}
|e(t)\rangle =A(t)|e\rangle +\sum_{n}a_{n}(t)|n\rangle ,
\end{equation}
the survival probability of the state $|e\rangle $, $s(t):=|\langle
e(0)|e(t)\rangle |^{2}=|A(t)|^{2}$, can be obtained without direct
measurements of the state $|e\rangle $. Actually, the measurements of the
states $|n\rangle $ provide us the probability $s(t)$ through the unitary
relation: 
\begin{equation}
s(t)=1-\sum_{n}|a_{n}(t)|^{2}.
\end{equation}
This operation is called as indirect measurement of $s(t)$. \ For the
specification of the indirect measurement, we should have to add one
important property. \ When we measure the states $|n\rangle $, we must
couple the states with some measurement devices. This measurement
interaction must commute with the original interaction in the system. \
Otherwise, the measurement \textit{directly} modifies the system dynamics
and the measurement cannot be regarded as \textit{purely indirect}.

In order to define generally and rigorously the purely indirect
measurements, let us define the core-zone and wave-zone states in the space $%
\mathcal{H}_{Z}$, in the same manner as \bigskip in the paper \cite{hm}.
\bigskip \newline
\bigskip {\huge Definition } \newline
We define the maximal subspace $\mathcal{H}_{W}$ ($\subset $ $\mathcal{H}%
_{Z} $), which is invariant under the advanced time evolution $U_{+}(t)=e^{-%
\frac{i}{\hbar }tH}$: $U_{+}(t)\mathcal{H}_{W}$ =$\mathcal{H}_{W}$ \ for all 
$t\geq 0$, where $H$ is the Hamiltonian of the system. \ In other words, \ \ 
\begin{equation}
If\quad |W\rangle \in \mathcal{H}_{W},\quad U_{+}(t)|W\rangle \in \mathcal{H}%
_{W}\quad \ for\quad all\quad t\geq 0.  \label{wz}
\end{equation}
This simply says that any vector $|W\rangle $ in $\mathcal{H}_{W}$ evolves
within $\mathcal{H}_{W}$ in the future. \ Then the space $\mathcal{H}_{Z}$
can be decomposed into mutually complementary subspaces $\mathcal{H}_{C}$
and $\mathcal{H}_{W}$:

\begin{equation}
\mathcal{H}_{Z}=\mathcal{H}_{C}\oplus \mathcal{H}_{W}.
\end{equation}
We call $\mathcal{H}_{C}$ the core subspace and $\mathcal{H}_{W}$ the
wave-zone subspace. \ For a practically meaningful systems, we assume both
are non-empty subspaces: \ 

\begin{equation}
\mathcal{H}_{C}\neq \o ,\ and\quad \mathcal{H}_{W}\neq \o .
\end{equation}
We call $|C\rangle (\in \mathcal{H}_{C})$ a core-zone state and $|W\rangle
(\in \mathcal{H}_{W})$ a wave-zone state.

To understand the physical meaning of the above definition of the wave-zone
state, let us consider that an unstable excitation in the core region of the
system decays and simultaneously emits photons, which propagate away from
the core and are measured by the outside devices. The wave-zone states
defined above faithfully describe the outside photon behaviors; the emitted
photons from the core eventually reach the wave zone, that is, the photons
begin to propagate freely in the space and never come back to the emission
core. Hence, advanced time evolution of the photon states in the wave zone
would never affect the transition amplitudes in the core-zone states. \ 

Clearly, the emergence of nontrivial subspace $\mathcal{H}_{W}$ is allowed
only when $\mathcal{H}_{W}$ is an open system. Also note that the invariance 
$U_{+}(t)\mathcal{H}_{W}$ =$\mathcal{H}_{W}$ \ itself cannot uniquely
determine the wave-zone subspace $\mathcal{H}_{W}$. \ We are interested in
the largest $\mathcal{H}_{W}$ and the smallest $\mathcal{H}_{C}$. \ The
boundary between $\mathcal{H}_{W} $ and $\mathcal{H}_{C}$ only admits a
one-way flow of states in the direction of future. \ 

Now we concentrate on the indirect measurements in the wave-zone subspace,
in which the wavefunction of the observed signals does not have
reflectional-wave contributions going back to the emission core after the
measurements. If a reflectional wave which goes back to the core exists and
produces the inverse decay process at the core, then this measurement cannot
be regarded as indirect. This is because the returning wave, after the
collision to the core, may excite the unstable state again. Therefore the
returning wave behaves just as an extra incident wave to the core system.
Hence, the reflectional effect, if it exists, ought to be called
``semi-direct'' and is out of our interest.

The above definition of the wave-zone subspace $\mathcal{H}_{W}$\ can also
be possible for the extended system which includes the measurement devices.
The non-existence of the reflectional wave induced by the measurement is
characterized by 
\begin{eqnarray}
P_{W}H_{m}P_{W} &=&H_{m},  \label{ewm} \\
P_{C}H_{i}P_{C} &=&H_{i}.
\end{eqnarray}
These equations leads $[H_{m},H_{i}]=0.$ \ In this paper, general \textit{%
purely indirect} measurements in the wave-zone subspace $\mathcal{H}_{W}$
are defined by this relation (\ref{ewm}).

Let us introduce the projection operator $P_{C}$ onto $\mathcal{H}_{C}$ as 
\begin{equation}
P_{C}=|e\rangle \langle e|+\sum_{c}|c\rangle \langle c|,
\end{equation}
where $\{|e\rangle ,|c\rangle \}$ is the complete orthonormal basis of $%
\mathcal{H}_{C}$. The projection operator onto $\mathcal{H}_{W}$ is 
\begin{equation}
P_{W}=\sum_{w}|w\rangle \langle w|,
\end{equation}
where $\{|w\rangle \}$ is the orthonormal complete basis of $\mathcal{H}_{W}$%
. The time evolution of the core-zone state $|e\rangle (\in \mathcal{H}_{C})$
now reads the following closed form; 
\begin{eqnarray}
|e(t)\rangle &=&U_{g+}(t)|e\rangle  \nonumber \\
&=&A(t)|e\rangle +\sum_{c}a_{c}(t)|c\rangle +\sum_{w}a_{w}(t)|w\rangle ,
\end{eqnarray}
where the operator $U_{g+}$ is the advanced time evolution operator
including the measurement effect which is defined by 
\begin{equation}
U_{g+}(t\geq 0):=\exp \left[ -i\frac{t}{\hbar }\left( H+gH_{m}\right) \right]
\end{equation}
and the normalization condition, $|A(t)|^{2}+\sum_{c}|a_{c}(t)|^{2}+%
\sum_{w}|a_{w}(t)|^{2}=1$, is satisfied.

The above wave-zone property eqn.(\ref{wz}) can be concisely described by
using the projection operators just introduced. \ The eqn.(\ref{wz}) means
``wave-zone states evolve within $\mathcal{H}_{W}$ in the future'':$%
[P_{W},U_{+}]P_{W}=0.$ Since $P_{C}+P_{W}=\mathbf{1}_{Z},$ this leads $%
[P_{C},U_{+}]P_{W}=0,$ which is equivalent to 
\begin{equation}
P_{C}U_{+}(t)P_{W}=0.  \label{c+w=0}
\end{equation}
Now let us define the survival probability of the genuine core-zone state $%
|e\rangle $ under the purely indirect measurement as 
\begin{equation}
s_{g}(t):=\left| \langle e|U_{g+}(t)|e\rangle \right| ^{2}.
\end{equation}
Then we prove a no-go theorem for the QZ paradox under the purely indirect
measurement as follows. \bigskip \newline
{\huge Theorem\bigskip } \newline
We prepare a system with a measurement apparatus, and suppose the total
state space admitts the decomposition into a core-zone subspace $\mathcal{H}%
_{C}$ and a wave-zone subspace $\mathcal{H}_{W}$. \ For this system, we
operate a purely indirect measurement with the Hamiltonian given by $gH_{m}$%
, where $g$ is a real coupling constant and $H_{m}$ is a Hermitian operator
satisfying $P_{W}H_{m}P_{W}=H_{m}$. Then, the survival probability $s_{g}(t)$
of a core-zone state $|e\rangle (\in \mathcal{H}_{C})$ does not depend on
the measurement coupling constant $g$; 
\begin{equation}
s_{g}(t):=\left| \langle e|U_{g+}(t)|e\rangle \right| ^{2}=\left| \langle
e|U_{+}(t)|e\rangle \right| ^{2}=s_{0}(t).  \label{th}
\end{equation}
\ \newline
This theorem implies that the measurement never affect the time evolution of
the state $|e(t)\rangle $ even if the measurement interaction becomes
extremely strong, $g\rightarrow \infty $. Consequently, the QZ paradox does
not take place under the purely indirect measurement.

The proof of the theorem is given as follows. Let us define a unitary
operator $V_{g}(t)$ as 
\begin{equation}
V_{g}(t)=U_{g+}(t)U_{+}^{\dagger }(t),
\end{equation}
where $V_{g}(0)=\mathbf{1}$ is satisfied. Then, this operator obeys the
following equation; 
\begin{equation}
i\hbar \partial _{t}V_{g}(t)=gV_{g}(t)U_{+}(t)H_{m}U_{+}^{\dagger }(t).
\end{equation}
By applying $P_{C}$ from the left in this equation and using the relation $%
H_{m}=P_{W}H_{m}P_{W}$, we obtain a differential equation of first order
with respect to time $t$; 
\begin{equation}
\partial _{t}[P_{C}V_{g}(t)]=-i\frac{g}{\hbar }[P_{C}V_{g}(t)][%
U_{+}(t)P_{W}H_{m}P_{W}U_{+}^{\dagger }(t)].  \label{dpc}
\end{equation}
Note that the initial condition for the operator $P_{C}V_{g}(t)$ is given by 
$P_{C}V_{g}(0)=P_{C}$. Under this initial condition, eqn(\ref{dpc}) is
solved at time $t\geq 0$ as 
\begin{equation}
P_{C}V_{g}(t\geq 0)=P_{C}.  \label{sol}
\end{equation}
Using the wave-zone property in the form eqn.(\ref{c+w=0}), this is easily
checked by substituting eqn(\ref{sol}) into eqn(\ref{dpc}). Rewriting eqn(%
\ref{sol}) yields the crucial relation; 
\begin{equation}
P_{C}U_{g+}(t)=P_{C}U_{+}(t).  \label{pcu}
\end{equation}
Now, using eqn(\ref{pcu}) and $\langle e|=\langle e|P_{C}$, we can
manipulate as follows. 
\begin{equation}
s_{g}(t)=\left| \langle e|P_{C}U_{g+}(t)|e\rangle \right| ^{2}=\left|
\langle e|P_{C}U_{+}(t)|e\rangle \right| ^{2}=s_{0}(t).
\end{equation}
Thus, the theorem is proved. Extension of the proof to the cases with many
coupling constants $g_{i}$ or coupling functions $g_{i}(x)$ is possible in
the similar way.

In the next section, we give a simple example to which the theorem can be
applicable rigorously.

\section{A Model}

\ In this section, we study a simple example which possesses a wave-zone
subspace and the no-go theorem for the QZ paradox is applicable rigorously.

The model is the same as that introduced in the previous paper \cite{hm}. In
what follows let us explain the model in detail. We set a two-level atom
system of size $d$ in one-dimensional space, in which $x$ denotes its
spatial coordinate, on the region $[-d/2,d/2]$. (Fig. 1 upper diagram) 
\begin{figure}[tbp]
\centerline{\includegraphics[width=84mm]{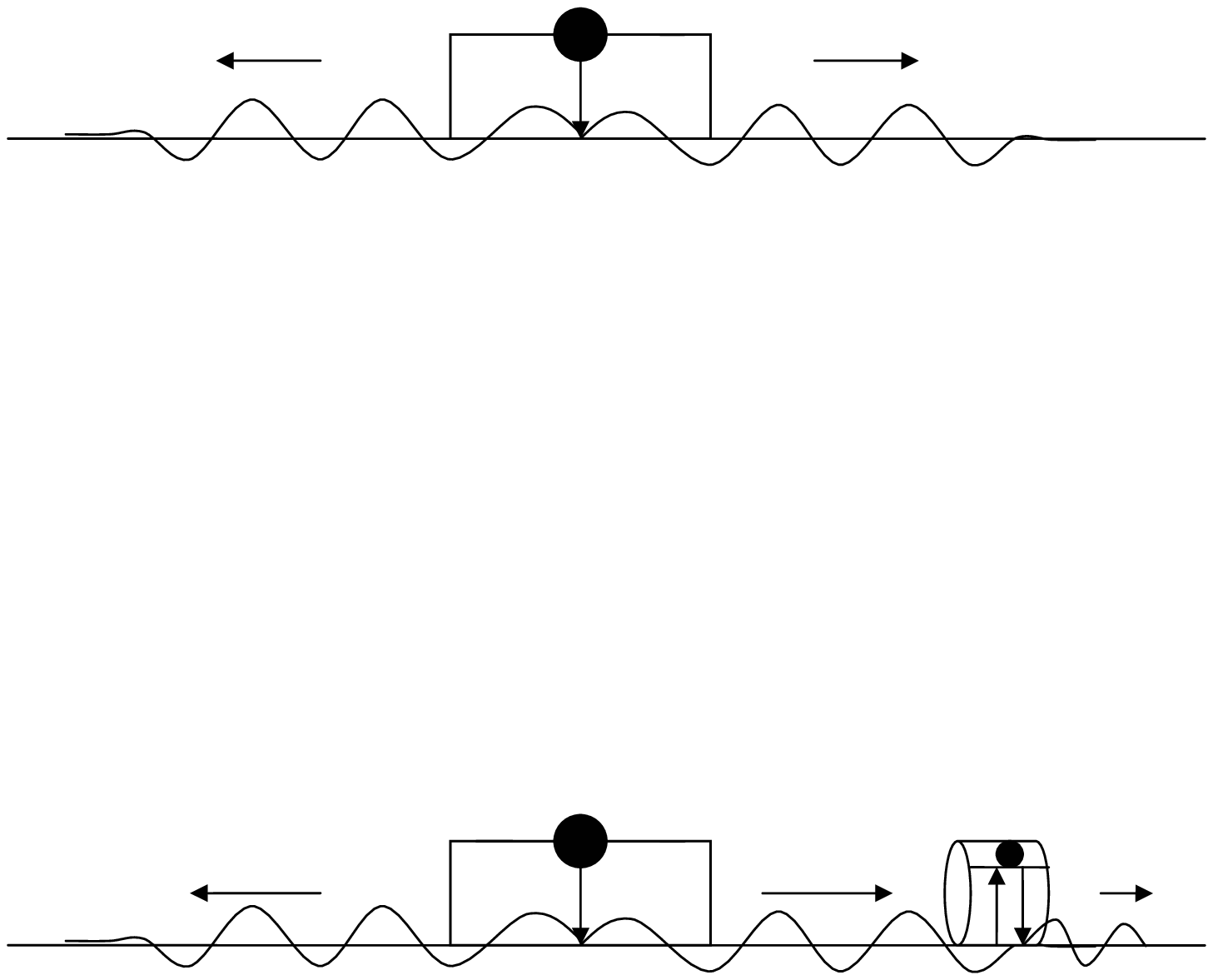}} \label{fig1}
\caption{ Schematic diagrams of the model. \newline
[Upper diagram] A two-level atom system of size $d$ in one-dimensional $x$%
-coordinate space. The system is located at $[-d/2,d/2]$, and emits spinor
pair when it decays. \newline
[Lower diagram] The system is coupled to a detector, which is located at $%
[x_{-},x_{+}]$ on the right hand side of the atom. The detector measures the
spinor field. }
\end{figure}
To express the upper and lower energy level states, we introduce a fermionic
pair of annihilation and creation operators $a$ and $a^{\dagger }$: 
\begin{equation}
\{a,\ a^{\dagger }\}=1,
\end{equation}
\begin{equation}
\{a^{\dagger },\ a^{\dagger }\}=\{a,\ a\}=0.
\end{equation}
We also introduce a massless spinor field 
\begin{equation}
\Phi (x)=(\Phi _{R}(x),\Phi _{L}(x))^{T}
\end{equation}
and quantize it in the fermionic way: 
\begin{equation}
\{\Phi _{h}(x),\ \Phi _{h^{\prime }}^{\dagger }(x^{\prime })\}=\delta
_{hh^{\prime }}\delta (x-x^{\prime }),
\end{equation}
\begin{equation}
\{\Phi _{h}^{\dagger }(x),\ \Phi _{h^{\prime }}^{\dagger }(x^{\prime
})\}=\{\Phi _{h}(x),\ \Phi _{h^{\prime }}(x^{\prime })\}=0,
\end{equation}
where $h(=R,L)$ is the helicity of the field excitations. As in the
Coleman-Hepp model \cite{ch} the vacuum state $|vac\rangle $ is introduced
using the annihilation operator as 
\begin{equation}
a|vac\rangle =0,
\end{equation}
\begin{equation}
\Phi _{h}(x)|vac\rangle =0.
\end{equation}
Then the excited state of the two-level atom is defined by 
\begin{equation}
|e\rangle =a^{\dagger }|vac\rangle .
\end{equation}
For the spinor field, we concentrate on the two particle states in which
only one R-helicity and one L-helicity particles exist. The state in which a
R-helicity particle stays at the position $x=x_{R}$ and a L-helicity
particle at $x=x_{L}$ is denoted by 
\begin{equation}
|x_{R},x_{L}\rangle =\Phi _{R}^{\dagger }(x_{R})\Phi _{L}^{\dagger
}(x_{L})|vac\rangle .
\end{equation}
Now let us write the Hamiltonian of the total system; it is composed of
three terms: 
\begin{equation}
H=H_{atom}+H_{\Phi }+H_{int}.  \label{totH}
\end{equation}
The first term $H_{atom}$ is the Hamiltonian of free motion of the two-level
atom and is given by 
\begin{equation}
H_{atom}=\hbar \omega a^{\dagger }a.
\end{equation}
The energy of the excited state is set to be $\hbar \omega $. The second
term $H_{\Phi }$ is the free Hamiltonian of the massless spinor field and is
defined by 
\begin{eqnarray}
H_{\Phi } &=&-i\hbar c\int_{-\infty }^{\infty }\Phi ^{\dagger }\sigma
_{3}\partial _{x}\Phi dx  \nonumber \\
&=&-i\hbar c\int_{-\infty }^{\infty }\left[ \Phi _{R}^{\dagger }\partial
_{x}\Phi _{R}-\Phi _{L}^{\dagger }\partial _{x}\Phi _{L}\right] dx,
\end{eqnarray}
where $\sigma _{3}$ is the third component of the Pauli matrix. If no
interaction term is added, the field Hamiltonian yields right-moving
particles for $h=R$ and left-moving particles for $h=L$ with the light
velocity. The third term $H_{int}$ in eqn.(\ref{totH}) expresses the
interaction between the two-level atom and the spinor field and is given by 
\begin{eqnarray}
H_{int} &=&\hbar \int_{-d/2}^{d/2}dx\int_{-d/2}^{d/2}dx^{\prime }  \nonumber
\\
&&\times \left[ g(x,x^{\prime })\Phi _{R}^{\dagger }(x)\Phi _{L}^{\dagger
}(x^{\prime })a+g(x,x^{\prime })^{\ast }a^{\dagger }\Phi _{L}(x^{\prime
})\Phi _{R}(x)\right] .  \label{16}
\end{eqnarray}
The interaction is supposed to take place only in the atom region $%
(-d/2,d/2) $, i.e. the support of $g(x,x^{\prime })$ is $(-d/2,d/2)$, and
the excited state of the atom decays into two particle states with different
helicities. Note that even after adding the interaction term in eqn.(\ref{16}%
), $|vac\rangle $ is still stable.

In this model the subspace, whose complete basis is given by $\{|e\rangle
,|x_{R},x_{L}\rangle \}$ with $x_{R}\geq -d/2$ and $x_{L}\leq d/2$, is
identified as $\mathcal{H}_{Z}$, because the evolution in this space is
closed: 
\begin{equation}
|\Psi (t)\rangle =C(t)|e\rangle +\int_{-d/2}^{\infty }dx_{R}\int_{-\infty
}^{d/2}dx_{L}F(x_{R},x_{L};t)|x_{R},x_{L}\rangle  \label{14}
\end{equation}
for an arbitrary vector $|\Psi \rangle (\in \mathcal{H}_{Z})$. From the
Schr\"{o}ddinger equation, the amplitudes obey the following equations: 
\begin{eqnarray}
&&i\partial _{t}C=\omega
C(t)+\int_{-d/2}^{d/2}dx_{R}\int_{-d/2}^{d/2}dx_{L}g(x_{R},x_{L})^{\ast
}F(x_{R},x_{L};t),  \label{1001} \\
&&(\partial _{t}+c\partial _{x_{R}}-c\partial
_{x_{L}})F(x_{R},x_{L};t)=-ig(x_{R},x_{L})C(t).  \label{1002}
\end{eqnarray}
This eqn.(\ref{1002}) can be integrated to yield 
\begin{eqnarray}
F(x_{R},x_{L};t) &=&F_{o}(x_{R}-ct,x_{L}+ct)  \nonumber \\
&&-i\int_{0}^{t}g(x_{R}-ct+c\tau ,x_{L}+ct-c\tau )C(\tau )d\tau ,
\label{1003}
\end{eqnarray}
where $F_{o}(x_{R},x_{L})$ is the initial amplitude of $F$ at $t=0$.

When $x_{R}>d/2$ or $x_{L}<-d/2$, by taking the initial condition as 
\begin{eqnarray}
&&C(0)=0,  \label{54} \\
&&F_{o}(x_{R},x_{L})=\delta (x_{R}-x_{R}^{\prime })\delta
(x_{L}-x_{L}^{\prime }),  \label{55}
\end{eqnarray}
the following relation arises from eqn.(\ref{1001}) and eqn.(\ref{1003}); 
\begin{equation}
U_{+}(t)|x_{R},x_{L}\rangle =|x_{R}+ct,x_{L}-ct\rangle .  \label{56}
\end{equation}
This means that the right- and left-moving particles propagate freely after
leaving the interaction region. It is worth stressing that even if only one
of the two conditions $x_{R}>d/2$ or $x_{L}<-d/2$ holds, the evolution in
eqn (\ref{56}) is still realized. This is because the interaction is
activated only when the both particles simultaneously stay in $(-d/2,d/2)$.

From the above result, we can introduce in $\mathcal{H}_{Z}$ a wave-zone
subspace $\mathcal{H}_{W}$ orthogonal to the excited state $|e\rangle $,
which is defined using the projection operator onto the subspace $P_{W}$: 
\begin{equation}
P_{W}=P_{R}+P_{L}+P_{RL},
\end{equation}
\begin{eqnarray}
P_{R} &=&\int_{d/2}^{\infty
}dx_{R}\int_{-d/2}^{d/2}dx_{L}|x_{R},x_{L}\rangle \langle x_{R},x_{L}|, \\
P_{L} &=&\int_{-d/2}^{d/2}dx_{R}\int_{-\infty
}^{-d/2}dx_{L}|x_{R},x_{L}\rangle \langle x_{R},x_{L}|, \\
P_{RL} &=&\int_{d/2}^{\infty }dx_{R}\int_{-\infty
}^{-d/2}dx_{L}|x_{R},x_{L}\rangle \langle x_{R},x_{L}|.
\end{eqnarray}
Then the core-zone subspace $\mathcal{H}_{C}$ is defined, using this
projection operator, as 
\begin{equation}
P_{C}:={}1_{Z}-P_{W}=|e\rangle \langle
e|+\int_{-d/2}^{d/2}dx_{R}\int_{-d/2}^{d/2}dx_{L}|x_{R},x_{L}\rangle \langle
x_{R},x_{L}|.  \label{pc}
\end{equation}
By construction, 
\begin{equation}
P_{C}|e\rangle =|e\rangle .
\end{equation}
Now the states are complete in $\mathcal{H}_{Z}$ as seen in eqn (\ref{14}),
and the wave-zone property in eqn(\ref{wz}) holds due to the relation eqn.(%
\ref{56}), we are ready to apply the theorem in the previous section. In
what follows, we discuss purely indirect measurements with finite errors in
the wave-zone subspace.

Let us consider that a measurement apparatus is set at the location $%
[x_{-},x_{+}]$ on the right hand side of the atom: $x_{+}>x_{-}>d/2$(Fig.1
lower diagram); the apparatus measures the $\Phi $ field with the following
interaction: 
\begin{eqnarray}
H_{m1} &=&\hbar \int_{-\infty }^{\infty }\Omega (k)b^{\dagger }(k)b(k)dk 
\nonumber \\
&&+\hbar \int_{x_{-}}^{x_{+}}dx\int_{-\infty }^{\infty }dk\lambda
_{R}(x,k)\left( b^{\dagger }(k)\Phi _{R}(x)+\Phi _{R}^{\dagger
}(x)b(k)\right) ,
\end{eqnarray}
where $b(k)$ ($b^{\dagger }(k)$) is the annihilation (creation) operator of
the detector excitations: 
\[
\lbrack b(k),\ b^{\dagger }(k^{\prime })]=\delta (k-k^{\prime }), 
\]
and $\Omega (k)$ and $\lambda _{R}(x)$ are real coupling functions. \ The
function $\lambda _{R}(x)$ has its support on $[x_{-},x_{+}]$ and controls
the incompleteness of the detector.

Here let us extend the projection operator $P_{W}$ to 
\begin{equation}
P_{W1}=P_{R}+P_{L}+P_{RL}+P_{ML},
\end{equation}
where 
\begin{equation}
P_{ML}:=\int_{-\infty }^{d/2}dx_{L}\int_{-\infty }^{\infty
}dk|k;x_{L}\rangle \langle k;x_{L}|,
\end{equation}
and 
\begin{equation}
|k;x_{L}\rangle :=b^{\dagger }(k)\Phi ^{\dagger }(x_{L})|vac\rangle .
\end{equation}
Then, it is checked that the subspace defined by $P_{W1}$ is really a
wave-zone subspace $P_{C}U_{+}(t)P_{W1}=0,$ complementary to $\mathcal{H}%
_{C} $; $P_{C}+P_{W1}=$ ${}1_{Z}$. Further it is trivial by construction
that 
\begin{equation}
P_{W1}H_{m1}P_{W1}=H_{m1}.
\end{equation}
Consequently we can apply the no-go theorem, and conclude that no QZ paradox
appears in this purely indirect measurement with finite errors.

It may be instructive to prove again the theorem for this model, explicitly
using the Schr{\"{o}}dinger equation. The time evolution in this case can be
expressed in a closed form as 
\begin{eqnarray}
|\Psi (t)\rangle &=&C(t)|e\rangle +\int_{-d/2}^{\infty }dx_{R}\int_{-\infty
}^{d/2}dx_{L}F(x_{R},x_{L};t)|x_{R},x_{L}\rangle  \nonumber \\
&&+\int_{-\infty }^{d/2}dx_{L}\int_{-\infty }^{\infty
}dkD_{k}(x_{L};t)|k;x_{L}\rangle .
\end{eqnarray}
The equations of motion read 
\[
i\partial _{t}C(t)=\omega
C(t)+\int_{-d/2}^{d/2}dx_{R}\int_{-d/2}^{d/2}dx_{L}g(x_{R},x_{L})^{\ast
}F(x_{R},x_{L};t), 
\]
\begin{eqnarray}
&&(\partial _{t}+c\partial _{x_{R}}-c\partial _{x_{L}})F(x_{R},x_{L};t) 
\nonumber \\
&=&-ig(x_{R},x_{L})C(t)-i\int_{-\infty }^{\infty }dk\lambda
_{R}(x_{R},k)D_{k}(x_{L};t),  \label{em2}
\end{eqnarray}
\[
i\partial _{t}D_{k}(x_{L};t)=\Omega
(k)D_{k}(x_{L};t)+\int_{x_{-}}^{x_{+}}dx_{R}\lambda
_{R}(x_{R},k)F(x_{R},x_{L};t). 
\]
It should be noticed here that the equations of motion for the core-zone
amplitudes $C(t)$ and $F(x_{R},x_{L})$ for $x_{R},x_{L}\in \lbrack -d/2,d/2]$
are not changed even after the measurement interaction is turned on.
Actually the eqn(\ref{em2}) exactly coincides with eqn.(\ref{1001}).
Further, since the coupling $\lambda _{R}$ vanishes in the region $%
[-d/2,d/2] $, eqn(\ref{em2}) is equivalent for the core-zone amplitudes to
eqn.(\ref{1002}). Therefore it is proven again that the finite couplings $%
\lambda _{R} $ and $\Omega (k)$, even if they are arbitrary large, give no
contribution to the survival probability. This is essentially because the
wave-zone amplitudes $F(x_{R},x_{L})$ for $(x_{R},x_{L})\in (-\infty ,\infty
)^{\otimes 2}\ominus \lbrack -d/2,d/2]^{\otimes 2}$ and $D_{k}(x_{L})$ are
unable to generate the core-zone amplitudes in the time evolution as
observed in the above equations. This implies that the wave-zone structure
is certainly maintained under the measurement and guarantees the no-go
theorem.

\section{Summary}

\ In this paper, we have investigated the resolution of the quantum Zeno
(QZ) paradox for purely indirect measurements with finite errors. Although
the QZ effect for direct measurements is theoretically proved and
experimentally demonstrated, the QZ paradox under the purely indirect
measurements cannot be realized without limitation. \ We have established a
natural condition under which no QZ effect is realized. \ This condition is
the wave-zone property described in eqn(\ref{wz}), which claims that the
information in the wave-zone states never flows into the core states in the
future. \ This one-sided property leads to the no-go theorem for QZ paradox
in eqn(\ref{th}). \ It is practically important that this property is not
destroyed by the introduction of a physical measurement apparatus which has
finite errors. \ 

By defining the survival probability of the system under the purely indirect
measurement with finite errors, we have found the probability is not
affected by the measurement at all. \ Further by using a simple model, we
have demonstrated the applicability of the general argument. \ Just watching
frequently from outside (observation) cannot be regarded as a measurement
which yields the quantum Zeno effect.

This type of wave-zone property seems to be quite common, especially in the
system which simultaneously includes microscopic and macroscopic components
such as Schr{\"{o}}dinger's cat system. \ From the view point of causality
in dynamics and measurement processes, the above condition wave-zone
property would yield general separation between microscopic and macroscopic
worlds. \ Many well known paradoxes associates with quantum mechanics seem
to be originated from the unlimited continuous concept of microscopic and
macroscopic objects. \ We hope our study in this paper can be a starting
point to resolves such prevailing paradoxes associated with quantum
mechanics. \ 

\ \newline
\textbf{Acknowledgement}\newline

We would like to thank A.Shimizu and K.Koshino for fruitful discussions.

\end{document}